# Numerical evaluation of Coulomb integrals for 1, 2 and 3-electron distance operators, $R_{C1}^{-n}R_{D1}^{-m}$, $R_{C1}^{-n}r_{12}^{-m}$ and $r_{12}^{-n}r_{13}^{-m}$ with real (n, m) and the Descartes product of 3 dimension common density functional numerical integration scheme

Sandor Kristyan

*Research Centre for Natural Sciences, Hungarian Academy of Sciences,*
*Institute of Materials and Environmental Chemistry*
*Magyar tudósok körútja 2, Budapest H-1117, Hungary,*

Corresponding author: kristyan.sandor@ttk.mta.hu

**Abstract**. Analytical solutions to integrals are far more useful than numeric, however, the former is not available in many cases. We evaluate integrals indicated in the title numerically that are necessary in some approaches in quantum chemistry. In the title, where R stands for nucleus-electron and r for electron-electron distances, the n, m= 0 case is trivial, the (n, m)= (1,0) or (0,1) cases are well known, a fundamental milestone in the integration and widely used in computational quantum chemistry, as well as analytical integration is possible if Gaussian functions are used. For the rest of the cases the analytical solutions are restricted, but worked out for some, e.g. for n, m= 0,1,2 with Gaussians. In this work we generalize the Becke-Lebedev-Voronoi 3 dimensions numerical integration scheme (commonly used in density functional theory) to 6 and 9 dimensions via Descartes product to evaluate integrals indicated in the title, and test it. This numerical recipe (up to Gaussian integrands with seed $\exp(-|\mathbf{r}_1|^2)$, as well as positive and negative real n and m values) is useful for manipulation with higher moments of inter-electronic distances, for example, in correlation calculations; more, our numerical scheme works for Slaterian type functions with seed $\exp(-|\mathbf{r}_1|)$ as well.
.

**Keywords.** Numerical evaluation of Coulomb integrals for one, two and three-electron distance operators,
Higher moment Coulomb distance operators $R_{C1}^{-n}R_{D1}^{-m}$, $R_{C1}^{-n}r_{12}^{-m}$ and $r_{12}^{-n}r_{13}^{-m}$ with real n, m≥0 and <0,
Generalization of 3 dimension Becke-Lebedev-Voronoi numerical integration scheme to 6 and 9 dimensions

## INTRODUCTION

Below, we use common notations, abbreviations and definitions: CC= correlation calculations; CQC= computational quantum chemistry; DFT= density functional theory; GTO= primitive Gaussian type atomic orbits, the $G_{Ai}$ in Eq.1; STO= primitive Slaterian type atomic orbits, that is, not $R_{Ai}^2 = |\mathbf{r}_i-\mathbf{R}_A|^2$ but $R_{Ai} \equiv |\mathbf{r}_i-\mathbf{R}_A|$ in Eq.1; HF-SCF= Hartree-Fock self consistent field; LC = linear combination; $\mathbf{R}_A \equiv (R_{Ax}, R_{Ay}, R_{Az})$ or $(x_A, y_A, z_A)$= 3 dimension position (spatial) vector of (fixed) nucleus A; $R_{AB} \equiv |\mathbf{R}_A-\mathbf{R}_B|$= nucleus-nucleus distance; $R_{Ai} \equiv |\mathbf{R}_A-\mathbf{r}_i|$= nucleus-electron distance; $\mathbf{r}_i \equiv (x_i,y_i,z_i)$= 3 dimension position (spatial) vector of (moving) electron i; $r_{ij} \equiv |\mathbf{r}_i-\mathbf{r}_j|$= electron-electron distance.

For one-electron density, $\rho(i) \equiv \rho(\mathbf{r}_i)$: Real$^3 \to$Real, the generally called one, two or three-electron Coulomb integrals $\int\rho(1)R_{C1}^{-n}R_{D1}^{-m}d\mathbf{r}_1$, $\int\rho(1)\rho(2)R_{C1}^{-n}r_{12}^{-m}d\mathbf{r}_1d\mathbf{r}_2$ and $\int\rho(1)\rho(2)\rho(3)r_{12}^{-n}r_{13}^{-m}d\mathbf{r}_1d\mathbf{r}_2d\mathbf{r}_3$ include the trivial case (n,m)=(0,0) reducing these to normalization $(\int\rho(1)d\mathbf{r}_1)^i= N^i$ for i=1,2 and 3, resp., where N is the number of electrons in the system, and the well known and most important cases, (n,m)=(1,0) or (0,1) yielding the integrals $\int\rho(1)R_{C1}^{-1}d\mathbf{r}_1$ and $\int\rho(1)\rho(2)r_{12}^{-1}d\mathbf{r}_1d\mathbf{r}_2$ for main energy values nuclear-electron attraction and electron-electron repulsion, resp., (strictly these are called Coulomb integrals), as well as if $\rho$ is approximated with LC of GTO, analytical evaluation is possible [1-3]. For the rest of values among n, m= 0,1,2 along with using GTO, analytical evaluations are also possible [4-6], but for general values (non-integer or higher or even negative n and m or using STO) analytic evaluations are not known, and numerical integration is necessary and provided below. These integrals provide smaller but (mathematically) important energy (correction) values in CC, e.g. in the so called R12 theory among many. As the analytical expressions are obviously different depending on (n, m) as well as the type (GTO, STO) used, the numerical integration scheme introduced is practically the same for integer, non-integer, positive or negative (n, m) as well as for GTO or STO. We test the numerical integration scheme for n, m=0,1,2 and GTO, where analytical integration is available [4-6], but the hypothesis is that the scheme works for all cases mentioned. Furthermore, integrals such as $\int \rho(1)\rho(2)\rho(3)\rho(4) r_{12}^{-n} r_{34}^{-m} d\mathbf{r}_1 d\mathbf{r}_2 d\mathbf{r}_3 d\mathbf{r}_4$ break up to simpler elements

$(\int \rho(1)\rho(2) r_{12}^{-n} d\mathbf{r}_1 d\mathbf{r}_2)(\int \rho(1)\rho(2) r_{12}^{-m} d\mathbf{r}_1 d\mathbf{r}_2)$, etc. and fall into the cases discussed.

The Coulomb energy between nuclei (index n) and electrons (index e) is calculated [1-3], for example, via DFT with integrals above using variable $\rho$ as $V_{ne} = N \Sigma_C Z_C \int \rho(1) R_{C1}^{-1} d\mathbf{r}_1$ (an exact expression [2, 7] if $\rho$ is exact) and $V_{ee} \approx (1/2) \int \rho(1)\rho(2) r_{12}^{-1} d\mathbf{r}_1 d\mathbf{r}_2$, while the HF-SCF theory uses one real (non-imaginary) Slater determinant, S, as $V_{ne} \approx N \Sigma_C Z_C \int S^2 R_{C1}^{-1}$ and $V_{ee} \approx (^N_2) \int S^2 r_{12}^{-1}$, where $\int$ is for N spin-orbit full space $d\mathbf{x}_1 \ldots d\mathbf{r}_N$., as well as similar expressions are used in configuration interactions methods [8]. (Notice that with W=1 from Eq.3: $(^N_2) \int S^*S d\mathbf{x}_1 \ldots d\mathbf{x}_N / (1/2) \int \rho(1)\rho(2) d\mathbf{r}_1 d\mathbf{r}_2 = (N-1)/N$.) The real (exact) or a physically realistic model $\rho(1) \geq 0$ (e.g. the HF level ground state $\rho(1) \equiv N \int S^2 ds_1 d\mathbf{x}_2 \ldots d\mathbf{r}_N$) can be approximated with an LC of GTO functions, using a well chosen GTO set $\{G_{Ai}\}$, where

$$G_{Ai}(a, nx, ny, nz) \equiv (x_i - R_{Ax})^{nx} (y_i - R_{Ay})^{ny} (z_i - R_{Az})^{nz} \exp(-a|\mathbf{r}_i - \mathbf{R}_A|^2) \quad (1)$$

are the primitive Gaussians with a>0 and nx, ny, nz $\geq 0$ benefiting its important property such as $G_{Ai}(a,nx,ny,nz)G_{Bi}(b,mx,my,mz)$ is also (an LC of) GTO; even more, derivatives of $G_{Ai}$ also preserves the form (breaking into an LC of GTO). The Coulomb interaction energies for molecular systems are expressed finally in LC of the famous elementary integrals $\int G_{A1} R_{C1}^{-1} d\mathbf{r}_1$ and $\int G_{A1} G_{B2} r_{12}^{-1} d\mathbf{r}_1 d\mathbf{r}_2$., i.e. the energy integrals above reduce to these, for which the analytic evaluation [1, 7, 9-10] has been fundamental and a milestone in the history of CQC. However, in CC (which estimates the error of HF-SCF via DFT, for example), one way [10] among the few famous and important ones [1-3] uses the integrals

$$\int G_{A1} G_{B2} G_{C3} W(1,2,3) \, d\mathbf{r}_1 d\mathbf{r}_2 d\mathbf{r}_3 \quad (2)$$

with weight

$$W(1,2,3) \equiv R_{C1}^{-n} R_{D1}^{-m}, \quad R_{C1}^{-n} r_{12}^{-m} \quad \text{or} \quad r_{12}^{-n} r_{13}^{-m} \quad (3)$$

as indicated in the main title.

Integrals in Eqs.2-3 belong mathematically to general manipulations in certain ways of CC, particularly to e.g. the so called "explicitly correlated R12 theories (ECT) of electron correlation", which bypass the slow convergence of conventional methods [1-2] by augmenting the traditional orbital expansions with a small number of terms that depend explicitly on the inter-electronic distance $r_{12}$. However, only approximate expressions are available for general evaluation, for example, Eq.52 in ref. [11] suggests for Eqs.2-3 that $\langle ijm|r_{12}^{-n} r_{13}^{-m}|kml\rangle \approx \Sigma_p \langle ij|r_{12}^{-n}|pm\rangle \langle pm|r_{12}^{-m}|kl\rangle$, where the bracket notation [1-2] is used along without reducing product GTO to single GTO, as well as the GTO basis set {p} for expansion has to be a "good quality" for adequate approximation; (notice that, particularly in ECT, for example, m=-n=1, because integrals with two negative powers in the distant operator do not occur). For particular n, m= 0,1,2 integers in Eqs.2-3, the mentioned refs. [4-6] provide analytical solutions for some other CC theories. Integrals in Eqs.2-3 would occur, e.g. in evaluations of matrix elements of the square of the Hamiltonian, such terms may occur when computing lower bounds to the energy. Furthermore, if derivatives appear, such as $\int (\partial \rho(\mathbf{r}_1)/\partial x_1)^p R_{C1}^{-n} d\mathbf{r}_1$, $\int (\partial \rho(\mathbf{r}_1)/\partial x_1)^p \rho(\mathbf{r}_2)^q r_{12}^{-n} d\mathbf{r}_1 d\mathbf{r}_2$ or many other algebraic possibilities frequently used in CC, and $\rho$ is given as LC of GTO, evaluation of integrals in Eq.2 are fundamental building blocks in calculations, since not only the products, but the derivatives of GTO in Eq.1 are also GTO. In summary, the numerical integration scheme introduced is useful and works for any positive or negative (but not extremely large absolute) real values of n, m.

## The extension of numerical integration scheme from 3 to 6 and 9 dimensions

The origin of the numerical integration scheme comes from Becke's recommendation for radial [12] and Lebedev's method for spherical [13-14] distribution, as well as the concept of Voronoi polygons to portion the molecular frame. For one-electron density, having shape as LC of GTO in Eq.1, this scheme falls into the simple form

$$\int f(\rho(1)) d\mathbf{r}_1 \approx \Sigma_{k=1,\ldots,L} c_k f(\rho(\mathbf{q}_k)), \quad (4)$$

where the set $\{c_k,\mathbf{q}_k\}$ is a specially chosen characteristic weight and coordinates in 3 dimensional space to estimate accurately the integral in the left hand side. Typical choices for $L=L_rL_s$ include $L_r=$ 20,...,200 radial points (ala Chebyshev) and $L_s$=20, 86, 302 spherical points (ala Lebedev), and since these are known [12-14], the procedure for numerical integration in Eq.4 is well defined, as well as widely tested in CC via DFT in CQC. The $\{c_k,\mathbf{q}_k\}$ is generated by a well defined algorithm [12-14] around all atoms (radially and spherically) in the molecular system.

The f in Eq.4 are special functions for CC [2-3, 7, 10] in current DFT methods, typically nonlinear in $\rho$, so analytical integration is generally not available. However, Coulomb related energies as $\int\rho(1)\rho(2)\rho(3)W(1,2,3)d\mathbf{r}_1d\mathbf{r}_2d\mathbf{r}_3 = \Sigma_g \int G_{A1}G_{B2}G_{C3}W(1,2,3)d\mathbf{r}_1d\mathbf{r}_2d\mathbf{r}_3$ in R12 theories for CC (e.g. Eq.3) or the total Coulomb energy ($W= R_{C1}$ or $r_{12}^{-1}$) itself is linear in $\rho(i)$, wherein $\rho$ can be well approximated in post-HF-SCF theory with LC of primitive GTO using that GTO times GTO are also GTO, ($\Sigma_g$ refers to this expansion including the coefficients), and it falls into terms in Eqs.2-3. But still, in this form, depending on W, analytical integrals are generally not available, except in some (n, m=0, 1, 2) cases [4-6, 9]. Even analytical form exists, the number of the GTO and computation steps in $\Sigma_g$ can be huge [15] depending on $\rho$ in spite of the many tricks [9, 15] used. However, in comparison to Eq.4 the $\Sigma_k$ contains the same number of terms independently of $\rho$, we say, the scheme introduced is practically the same for all W: The main idea of this work is the generalization of Eq.4 and test

$$\int f(\rho(1),\rho(2),\rho(3))d\mathbf{r}_1d\mathbf{r}_2d\mathbf{r}_3 \approx \Sigma_{A,B,C}\, c_Ac_Bc_C\, f(\rho(\mathbf{q}_A),\rho(\mathbf{q}_B),\rho(\mathbf{q}_C)), \quad (5)$$

particularly, for

$$f= \rho(1)\rho(2)\rho(3)W(1,2,3). \quad (6)$$

If $W= W(1)W(2)W(3)$, most simply $W=1$ or e.g. $W=R_{D1}^{-3}R_{E2}^4$, etc., Eq.5 reduces to product of three ones in Eq.4. Eq.5 is the extension of Eq.4 from 3 to 9 dimensions, extension to 6 dimension case ($\int f(\rho(1)\rho(2))d\mathbf{r}_1d\mathbf{r}_2$) is obviously a back reduction of Eq.5 to sum $\Sigma_{A,B}$.

Important is that set $\{c_k,\mathbf{q}_k\}$ in Eq.4 is the same kind (or exactly the same) in Eq.5, but combining each element to each as triplets, coming from the fact that $\rho(i)$ is the same 3 dimensional one-electron density for the 9 dimension integral in Eq.5. Simply, A, B and C run for the same 1,2,…,L points, so Eq.5 contains $L^2$ terms (for case $f(\rho(1)\rho(2))$) or $L^3$ (for case $f(\rho(1)\rho(2)\rho(3))$) to calculate. There are two reasons to choose different point sets $\{c_k,\mathbf{q}_k\}$ for the two (i=1,2) or three (i=1,2,3) variables $\mathbf{r}_i$: 1.: A choice is to reduce the computation task, that is, if $L=L_sL_r$ is not the same, for example for $\mathbf{r}_1$ and $\mathbf{r}_2$, the larger point set (larger mesh) provides finer numerical integration for that variable in the integrand in Eq.5 if necessary, determined by f. 2.: Another reason is a must: Important technical issue is that if W contains reciprocal distance, e.g. $r_{12}^{-n}$ with n>0, as a frequent case, the $r_{12}=|\mathbf{q}_A-\mathbf{q}_B|=0$ for some A=B indexes in the sum blows the reciprocal value up; (which is not problem if n≤0). In this case, two different kinds of point sets must be chosen. It means that, around an atom the same spherical mesh (ala Lebedev) is picked up in slightly different radius (ala Chebyshev) for $\mathbf{r}_1$ vs. $\mathbf{r}_2$, or two different spherical meshes with the same radius: Technically, the $L_s$ value of spherical (or angular) points is picked automatically (from 20, 86 or 302) by the atom types (atomic number Z) as suggested by ref.[16] based on practical reasons in CQC, (although in principle it can be an independent choice, as well as $L_r$ can also be made automatic [16]), so one can tackle with the value of $L_r$ radial points, e.g. $L=L_s$(automatic)$L_r$(=20…200) can be chosen as $L_1$= 200$L_s$(automatic) and $L_2$= 190$L_s$(automatic), to generate the sets as $\{c^{L1}_A,\mathbf{q}^{L1}_A\}$ for $\mathbf{r}_1$ and $\{c^{L2}_B,\mathbf{q}^{L2}_B\}$ for $\mathbf{r}_2$, so possibly $r_{12}\neq 0$ for any (A=B and A≠B) spatial point pair; if accidentally the overlap set is still not empty, that is $\{c^{L1}_A,\mathbf{q}^{L1}_A\}\cap\{c^{L2}_B,\mathbf{q}^{L2}_B\}\neq 0$, one can tune $L_r$ to 180, etc.; keeping in mind that larger $L_r$ yields slightly more accurate numerical integration.

We make a note on a first correction or factorization in relation to the important Eq.6 to improve the accuracy. Eqs.5 and 6 yield

$$\int\rho(1)\rho(2)\rho(3)W(1,2,3)d\mathbf{r}_1d\mathbf{r}_2d\mathbf{r}_3 \approx \Sigma_{A,B,C}\, c_Ac_Bc_C\, \rho(\mathbf{q}_A)\rho(\mathbf{q}_B)\rho(\mathbf{q}_C)\, W(\mathbf{q}_A,\mathbf{q}_B,\mathbf{q}_C) \quad (7)$$

for the main title of this work. (If e.g. $W=r_{12}^4 r_{13}^5$ in Eq.3, then in Eqs.5-7 it means $W=|\mathbf{q}_A-\mathbf{q}_B|^4|\mathbf{q}_A-\mathbf{q}_C|^5$.) If W=1, Eq.7 reduces to $N^3= (\int\rho(1)d\mathbf{r}_1)^3 = \int\rho(1)\rho(2)\rho(3)d\mathbf{r}_1d\mathbf{r}_2d\mathbf{r}_3 \approx \Sigma_{A,B,C}\, c_Ac_Bc_C\, \rho(\mathbf{q}_A)\rho(\mathbf{q}_B)\rho(\mathbf{q}_C)=$ $(\Sigma_A\, c_A\, \rho(\mathbf{q}_A))^3 = (0.995\, N)^3= 0.985\, N^3$, if the same mesh is used, as well as if we have in the far right that the numerical integral makes, let say, 1-0.995= 0.005= 0.5% error via Eq.4, and as a consequence, the final error increases to 1-0.985= 1.5% error via Eq.5. Dividing these two equation yields

$$\int\rho(1)\rho(2)\rho(3)W(1,2,3)d\mathbf{r}_1d\mathbf{r}_2d\mathbf{r}_3\approx$$
$$[\int\rho(1)d\mathbf{r}_1/\Sigma_{A}c_A\rho(\mathbf{q}_A)]^3\, \Sigma_{A,B,C}c_Ac_Bc_C\rho(\mathbf{q}_A)\rho(\mathbf{q}_B)\rho(\mathbf{q}_C)W(\mathbf{q}_A,\mathbf{q}_B,\mathbf{q}_C). \quad (8)$$

For the right side, the $\int\rho(1)d\mathbf{r}_1$ can be evaluated analytically in practice, because it is LC of GTO in Eq.1, more it is generally normalized to N. In the square bracket the ratio is 1 if Eq.4 is 100% accurate, (as well as if W=1 in Eq.8, it becomes trivially equal). One must keep in mind that Eq.8 supposes that same mesh is used for all three variables. If three different meshes are chosen (see above), more, three different densities (distinguished by indexes $\alpha,\beta,\gamma$ as in the test Table 1 below), the factor is

$$[ \int\rho_\alpha(1)d\mathbf{r}_1 \int\rho_\beta(1)d\mathbf{r}_1 \int\rho_\gamma(1)d\mathbf{r}_1 /( \Sigma_A c_A \rho_\alpha(\mathbf{q}_A) \Sigma_B c_B \rho_\beta(\mathbf{q}_B) \Sigma_C c_C \rho_\gamma(\mathbf{q}_C) )], \quad (9)$$

where the indexes A, B and C represent the three (generally different) meshes for variables $\mathbf{r}_1$, $\mathbf{r}_2$ and $\mathbf{r}_3$. Since the triple sum increases the number of points to calculate, see $L^3$ above, this factorization may allow reducing the value for L (or particularly for one or two among $L_1$, $L_2$ and $L_3$). The reduction of 9 dimension (Eqs.8-9) to 3 ($\int\rho(1)W(1)d\mathbf{r}_1$) and 6 ($\int\rho(1)\rho(2)W(1,2)d\mathbf{r}_1d\mathbf{r}_2$) dimension is straightforward.

Jumping from 3 (Eq.4) to 6 and 9 dimensions (Eqs.5-6) increases the time of computation (L to $L^2$ and $L^3$ terms in sum). An approximate reduction from 9 to 6 is as follow: $\Sigma_A\Sigma_B f(A)g(B) = \Sigma_A f(A)\Sigma_B g(B)$ cannot be done if f and g are coupled via a W, so e.g. approximation in ref.[11] can be formulated here as e.g. $\int\rho(1)\rho(2)\rho(3) r_{12}^n r_{13}^m d\mathbf{r}_1 d\mathbf{r}_2 d\mathbf{r}_3 \approx \int F(\rho(1))\rho(2)r_{12}^n d\mathbf{r}_1 d\mathbf{r}_2 \int F(\rho(1))\rho(3)r_{13}^m d\mathbf{r}_1 d\mathbf{r}_3$, etc., where F should be wisely chosen (e.g. F= $\rho^{1/2}$), a rougher approximation than the relatively accurate mesh $\{c_k,\mathbf{q}_k\}$ which is used for the product in right side, but the computation is not to sum up $L^3$ terms but only $2L^2$.

## Test of the numerical integration scheme for 6 and 9 dimensions

We test the accuracy of Eqs.5-8 for 1.: Elementary cases in Eqs.2-3, where analytical integral formulas are known [4-6, 9]. This is a very strong test in view of Eq.6, because in that case f is linear in $\rho(i)$, and $\rho$ is also about LC of the primitive GTO in Eq.1, so the left hand sides of Eqs.5-8 reduce to Eq.2 in accuracy test. 2.: HF-SCF/basis optimized Slater determinant, $S=S_0$, and ground state one-electron density, $\rho=\rho_{0,HF}$, yields $V_{ee}^{HF} \equiv (N(N-1)/2)\int S^*S\ r_{12}^{-1} \approx V_{ee}^{DFT,main\ term} \equiv (1/2)\int\rho(1)\rho(2)r_{12}^{-1}d\mathbf{r}_1 d\mathbf{r}_2$, both approximates the true electron-electron repulsion energy with correlation and basis set error, but in principle they are not equal; the analytical evaluation of the former correlates to the numerical evaluation of the latter. Notice that with normalization properties and without weight $W=r_{12}^{-1}$, the ratio $(N(N-1)/2)\int S^*S\ /\ (1/2)\int\rho(1)\rho(2)d\mathbf{r}_1d\mathbf{r}_2 = (N(N-1)/2)/((1/2)N^2) = (N-1)/N \rightarrow 1$ if $N \rightarrow \infty$. One can test for value (N-1)/N with $V_{ee}^{HF}$ (with S from HF-SCF/STO-3G and analytical Eq.2) vs. $V_{ee}^{DFT,main\ term}$ (with $\rho$ from HF-SCF/STO-3G and numerical Eqs.5-8), both with $W=r_{12}^{-1}$. Positive tests validate the scheme or procedure in Eq.5 for the main title of this work.

Numerical integration scheme in Eq.4 is widely used in DFT integration. When $f(\rho(1))$ is (LC of) primitive GTO (Eq.1), and analytical integral is available for the left side of Eq.4, the numerical integration (right side of Eq.4) follows it up to many decimal digits, and the hypothesis is that if $f(\rho(1))$ functions are nonlinear functions of $\rho$ and analytic integration is not available, the Eq.4 still works, as experienced in CQC for CC. In our case, our hypothesis is the same for Eq.5, even though we have tested Eq.6 only.

Another extension is that, 1.: If not GTO, but STO is used, the analytical evaluation is far more difficult for Eq.2, almost impossible, except for very elementary cases. A simple escape route is to use the approximation $\exp(-p|\mathbf{r}_1-\mathbf{R}_P|) \approx \Sigma_{(i)} c_i G_{P1}(a_i,0,0,0)$, which is well known in molecular structure calculations, see the idea of STO-3G basis sets [3] and higher levels in which one does not even need many terms in the summation, so GTO are evaluated analytically if possible instead of STO, but in fact in this way, one loses the full analytical evaluation of the original task, however for full numerical integration this is obviously unnecessary. In summary, focusing on the practically important Eqs.3, 5 and 6: a.: If $\rho$ is LC of STO, then analytical integration is almost impossible, b.: If $\rho$ is LC of GTO, then analytical integration is available for integer n, m=0,1,2 [4-6, 9], c.: If $\rho$ is LC of STO or GTO, then the numerical scheme in Eqs.7-9 works for any positive or negative real value n and m, (of course, for very large |n|, |m| >> 2-4 values, the radial and spherical schemes reported in refs.[12-14] are not high level enough and needs to be evaluated for). 2.: For our numerical integration the n and m can be negative in Eq.3 as well. (Notice that, if $W= r_{12}^{-n}$ in Eq.3 $\Rightarrow$ Eq.2 factorizes to three for easy analytical evaluation if n=-2, but do not if n=-1, for example, etc.)

Eqs.7-9 work if $\rho$ is an LC (more, a product of LC's, since STO does not preserve the form of product) of either GTO or STO. We make a note on the opportunity to use STO via Eqs.7-8 in post HF-SCF calculations as alternative to GTO. (In HF-SCF, 1.: The $W= R_{C1}^{-1}$ and $r_{12}^{-1}$ multipliers and differential operators ($\nabla$, $\partial/\partial x_1$) come up only in the integrals, 2.: In CQC the STO is more realistic than GTO, 3.: GTO is used because GTO preserves the form of products, but STO does not, so analytical integration is possible with GTO. However, with Eqs.7-9 all the HF-SCF integrals can be evaluated numerically for both, STO and GTO. 4.: In the concept of e.g. STO-3G, 6-31G**, etc. basis sets, the STO is approximated with 3, 6, etc. GTO.) If all analytical integrals are replaced with numerical via the 3 and 6 dimension versions of Eqs.7-9, smaller basis set could be enough. It provides fewer LCAO parameters and smaller Hamiltonian matrix, which reduces the computation time and increases the computation stability, finally, allowing for calculating larger molecular systems.

In the test we have chosen the magnitude of GTO exponents from STO-3G basis set of hydrogen (H, 3.4252509D+00, 6.2391373D-01, 1.6885540D-01) and carbon (C, 7.1616837D+01, 1.3045096D+01, 3.5305122D+00) atoms, that is, p, q:=0.3, 1, 3, 15, 70 with $R_{PQ}$:= bond (1 A), van der Waals (4 A) and far away (10 A) magnitude lengths ($\approx$ 2, 8, 20 bohr) and $\mathbf{R}_P$:=(0,0,0). The STO exponents have come from atomic orbits containing exp(-Z|$\mathbf{r}_1$|/n) with Z=1, 6 for H and C with quantum numbers n=1-3, i.e. p, q:= 0.3, 1, 3, 6 scans Z/n. Boys functions [4-6]: $F_L(v) \equiv \int_{(0,1)} \exp(-vw^2)w^{2L}dw$, L=0,1,2,..., v>0 or v≤0 and $2vF_{L+1}(v) = (2L+1)F_L(v) - \exp(-v)$. High ($L_r$=100, 102) medium (50, 52, 54) and low (20, 22, 24) radial meshes were tested for ($\mathbf{r}_1$, $\mathbf{r}_2$, $\mathbf{r}_3$) on a 1400.000 MHz CPU machine.

**TABLE 1.** Comparison of analytical and numerical integration

| 100(1-(numerical/analytical))= % error by Eqs.7-9 using any of $\mathbf{R}_Q$= (d,0,0) and $\mathbf{R}_S$= (0,d,0) with d= 0, 2, 8, 20, and any of p, q, s = 0.3, 1, 3, 15, 70 for GTO or any of p, q = 0.3, 1, 3, 6 for STO. | \|% error\| by Eq.7: $L_r$=20, 22, 24 $L_r$=50, 52, 54 $L_r$=100, 102 Integral value | \|% error\| by Eqs.8-9: $L_r$= 20, 22, 24 $L_r$=50, 52, 54 $L_r$=100, 102 |
|---|---|---|
| **Trivial case:** product of primitive spherical GTO and W=1 $\int_{(R6)} \exp(-p\ R_{P1}^2) \exp(-q\ R_{Q2}^2) d\mathbf{r}_1 d\mathbf{r}_2 =$ $\int_{(R3)} \exp(-p\ R_{P1}^2)d\mathbf{r}_1 \int_{(R3)} \exp(-q\ R_{Q2}^2)d\mathbf{r}_2 =$ $\pi^3/(pq)^{3/2}$ (=$\pi^3/p^3$ if p=q) vs. Eqs.7-9. | $10^{-1}$-$10^{-4}$ $10^{-6}$-$10^{-9}$ $10^{-6}$-$10^{-10}$ $10^3$-$10^{-4}$ | 0.0 (trivial) 0.0 (trivial) 0.0 (trivial) |
| **Trivial case:** product of primitive spherical STO and W=1 $\int_{(R6)} \exp(-p\ R_{P1}) \exp(-q\ R_{Q2}) d\mathbf{r}_1 d\mathbf{r}_2 =$ $\int_{(R3)} \exp(-p\ R_{P1})d\mathbf{r}_1 \int_{(R3)} \exp(-q\ R_{Q2})d\mathbf{r}_2 =$ $64\pi^2/(pq)^3$ (=$64\pi^2/p^6$ if p=q) vs. Eqs.7-9. | 1-$10^{-5}$ $10^{-3}$-$10^{-9}$ $10^{-4}$-$10^{-9}$ $10^6$-$10^{-9}$ | 0.0 (trivial) 0.0 (trivial) 0.0 (trivial) |
| **Elementary case:** one-electron spherical GTO $\int_{(R3)} \exp(-p\ R_{P1}^2) R_{Q1}^{-1} d\mathbf{r}_1 = (2\pi/p) \int_{(0,1)} \exp(-p\ R_{QP}^2 w^2)dw = (2\pi/p)F_0(v)$, v≡ p $R_{QP}^2$, [9] vs. Eqs.7-9. | 1-$10^{-4}$ 1-$10^{-7}$ $10^{-2}$-$10^{-8}$ 20-5*$10^{-4}$ | 1-$10^{-8}$ 1-$10^{-10}$ $10^{-2}$-$10^{-10}$ |
| **Elementary case:** one-electron spherical GTO $\int_{(R3)} \exp(-p\ R_{P1}^2) R_{Q1}^{-2} d\mathbf{r}_1 = (2\pi^{3/2}/p^{1/2}) \int_{(0,1)} \exp(p\ R_{QP}^2 (w^2-1))dw =$ $(2\pi^{3/2}/p^{1/2})e^{-v}F_0(-v)$, v≡p $R_{QP}^2$,[4-6] vs. Eqs.7-9. | 0.5-$10^{-2}$ 0.1-$10^{-3}$ 0.1-$10^{-3}$ 20-2*$10^{-5}$ | 0.5-$10^{-3}$ 0.1-$10^{-3}$ 0.1-$10^{-3}$ |
| **Elementary case:** two-electron spherical GTO $\int_{(R6)} \exp(-p\ R_{P1}^2) \exp(-q\ R_{Q2}^2) r_{12}^{-1} d\mathbf{r}_1 d\mathbf{r}_2 =$ $(2\pi^{5/2}/(pq)) \int_{(0,c)} \exp(-pqR_{PQ}^2 w^2)dw$, c≡$(p+q)^{-1/2}$, [9] vs. Eqs.7-9. | 4-$10^{-4}$ 4-$10^{-8}$ 4-$10^{-9}$ 5*$10^2$-5*$10^{-6}$ | 4-$10^{-10}$ 4-$10^{-10}$ 4-$10^{-9}$ |
| **Elementary case:** two-electron spherical GTO $\int_{(R6)} \exp(-p\ R_{P1}^2) \exp(-q\ R_{Q2}^2) r_{12}^{-2} d\mathbf{r}_1 d\mathbf{r}_2 =$ $2\pi^3(pq)^{-1/2}(p+q)^{-1}\int_{(0,1)} \exp(v(w^2-1))dw = (2\pi^3(pq)^{-1/2}(p+q)^{-1})e^{-v}F_0(-v)$, v≡ $pqR_{PQ}^2/(p+q)$, [4-6] vs. Eqs.7-9. | 5-$10^{-5}$ 5-$10^{-6}$ 5-$10^{-6}$ 5*$10^2$-$10^{-5}$ | 5-$10^{-6}$ 5-$10^{-6}$ 5-$10^{-6}$ |
| **Elementary case:** two-electron spherical GTO $\int_{(R6)} \exp(-pR_{P1}^2) \exp(-qR_{Q2}^2) R_{S1}^{-1}r_{12}^{-1}d\mathbf{r}_1 d\mathbf{r}_2 =$ $(4\pi^2/q)\int_{(0,1)}F_0(\|p\mathbf{R}_P+qu^2\mathbf{R}_Q-g\mathbf{R}_S\|^2)g^{-1}\exp(-f/g)du$, f≡ $pqR_{PQ}^2u^2$ and g≡ $p+qu^2$, [4-6] vs. Eqs.7-9. | 5-0.01 5-0.001 5-0.001 5*$10^2$-$10^{-6}$ | 5-0.01 5-0.001 5-0.001 |
| **Elementary case:** three-electron spherical GTO $\int_{(R9)} \exp(-p\ R_{P1}^2) \exp(-q\ R_{Q2}^2) \exp(-s\ R_{S3}^2)r_{12}^{-1}r_{13}^{-1} d\mathbf{r}_1 d\mathbf{r}_2 d\mathbf{r}_3 =$ $(4\pi^{7/2}/(qs))\int_{(0,1)}\int_{(0,1)} g^{-3/2}\exp(-f/g)dudt$, f≡ $pqR_{PQ}^2u^2+psR_{PS}^2t^2+qsR_{QS}^2u^2t^2$, g≡ $p+qu^2+st^2$, [4-6] vs. Eqs.7-9. | 1-$10^{-4}$ - - 8*$10^3$-$10^{-9}$ | 1-$10^{-5}$ - - |

Conclusion from the Table 1: Few radial points ($L_r=20$) compete with larger ones in numerical integration, good for reducing computation time, but notice that, the analytical expression is one equation with a few operations in contrast to $(L_rL_s)^m$ in numerical integration by Eqs.4-9 ($m=1,2,3$). The W in Eq.3 can deform the angular and radial symmetry if $W\neq1$ in ways that may decrease the numerical accuracy, however, the situation is the same in general CC via Eq.4. Most importantly, the numerical integration scheme above is acceptable until analytical integration is not available and recommended e.g. in CC. Figure 1 shows some intermediate values for exponents in Eq.3.

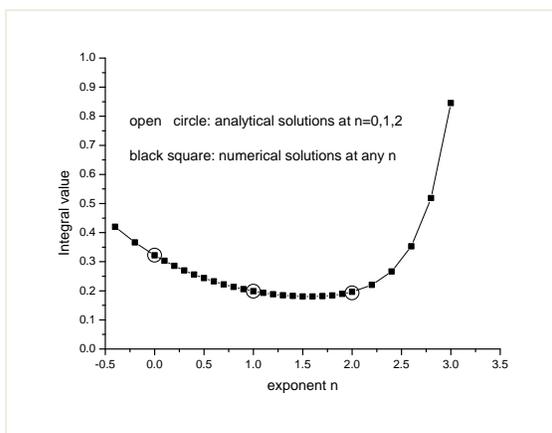

**FIGURE 1.** From the 1st, 5th and 6th lines of the Table 1 with $R_{PQ}=0$ ($\Rightarrow F_0(0)=1$), the analytical expressions in the range $n\in[0,2]$ are
$\int_{(R6)} \exp(-p\, R_{P1}^2)\exp(-q\, R_{P2}^2)\, r_{12}^{-n}\, d\mathbf{r}_1 d\mathbf{r}_2 =$
$\pi^3(pq)^{-3/2}$ if $n=0$, $2\pi^{5/2}(pq)^{-1}(p+q)^{-1/2}$ if $n=1$ and $2\pi^3(pq)^{-1/2}(p+q)^{-1}$ if $n=2$. Other values can be calculated numerically with Eq.7 as plotted with, let say, $(p, q)=(0.3, 70)$ and $L_r = 20, 22$.

## ACKNOWLEDGMENTS

Financial and emotional support for this research from OTKA-K- 2015-115733 and 2016-119358 are kindly acknowledged. The subject has been presented in ICNAAM 2018_40, Greece, Rhodes.